# Spin Relaxation in Silicon Nanowires

Ashish Kumar[a)] and Bahniman Ghosh

*Department of Electrical Engineering, Indian Institute of Technology, Kanpur 208016, India*

Abstract - We simulate spin polarized transport of electrons along a silicon nanowire and along a silicon two dimensional channel. Spin density matrix calculations are used along with the semi-classical Monte Carlo approach to model spin evolution along the channel. Spin dephasing in silicon is caused due to Rashba Spin Orbit Interaction (structural inversion asymmetry) which gives rise to D'yakonov-Perel' relaxation. Spin relaxation length in a nanowire is found to be higher than that in a 2-D channel. The effect of driving electric field on spin relaxation is also investigated. These results obtained are essential for design of spintronics based devices.

a). Electronic mail : ashish12.kumar@gmail.com

# I. INTRODUCTION

The study in the area of spintronics has gained momentum over years due to the potential the spintronics based devices possess [1]. While much breakthrough has been achieved in metal based spintronics [2], semiconductor based spintronics is still in a developing stage. The integration of contemporary electronics with spintronics will result in devices that can perform much more [2-5] than what is possible with the contemporary semiconductor technology. Spintronics based devices would require lesser power than contemporary semiconductor electronics and would be much faster. The transmission would be dissipation less leading to smaller devices. Storing information into spins of electrons will lead to non-volatile memories. The plethora of advantages that spintronics provides leads to an ongoing quest to ascertain semiconductor materials that are suitable for use in spintronics devices.

Spin injection at the source, spin transport through the material and spin detection at the drain are the three main processes that a spintronic device works on. Of the three processes, our paper deals with the second process of spin transport. Spin relaxation is an integral process in the study of spin transport. Spin relaxation lengths signify the distance from the source in which the spin information of the electrons gets lost. Hence spin relaxation lengths are critical when deciding upon the suitability of a material for spintronics.

Much theoretical and experimental work has been done to investigate spin relaxation in semiconductors. Various III-V and II-VI compounds have been studied [5-9]. In Ref.[7] spin transport in GaAs is studied experimentally using a spectroscopic method. In Ref. [8] Monte Carlo method is used to simulate spin polarized transport in GaAs/GaAlAs quantum wells. Spin relaxation is reinvestigated at different conditions and for different dimensionality of systems [13,14]. In Ref.[13] Monte Carlo method is used to model spin transport in GaAs nanowires. In Ref.[14] spin polarized transports in 1D and 2D III-V heterostructures are compared. Much experimental work has been done with germanium [10] and silicon [11,12] as well. However theoretical work on spin relaxation in silicon, that forms the backbone of semiconductor devices, is still in its very early stages and thus remains poorly comprehended with regards to its spin transport properties. Suitability of silicon as a spintronic material, if established, will lead to seamless integration of semiconductor technology with spintronics and this motivates us to investigate spin relaxation properties of silicon. To the best of our knowledge, Monte Carlo simulations to study spin transport in silicon have still not been reported and this leads us to undertake spin relaxation studies on silicon which still remains a very potent semiconductor material.

In this paper, we investigate spin relaxation in 2-D silicon channels and 1-D silicon nanowires and draw a comparison between them. Silicon nanowires have attracted great attention due to their potential to perform multiple functions [15-17]. Silicon nanowires find use in producing efficient thermoelectric devices [15]. They have the potential to function as logic devices [16] and as biological sensors. This necessitates assessing their performance for application in spintronics.

Of the three spin relaxation mechanisms, spin dynamics in silicon systems is dominated by D'yakonov-Perel (DP) [18] mechanism. Bir-Aronov-Pikus [20] mechanism is present in p type semiconductors only. Electron spin flip [21] due to Elliot Yafet (EY) [19] mechanism is also almost absent in silicon since it is a larger bandgap material with very small spin-orbit splitting ($E_{so}$=0.044eV).

We use semiclassical Monte Carlo approach to model electron transport in 2D channels and in 1D silicon nanowires. The Monte Carlo method [22,23,24] coupled with spin density matrix [24] dynamics models the spin transport of electrons in both 2-D and 1-D systems. As in some of the previous works done [13], improvement in spin relaxation on confinement is observed. Spin relaxation is investigated at different applied electric fields.

## II. MODEL

A comprehensive account of the Monte Carlo method [22,23,24] and spin transport model [6,24] is presented elsewhere. We shall restrict our discussion only to the essential features of the model. According to the co-ordinate system chosen, *x* is along the length of the device, *y* is along the width of the device and *z* is the along the thickness of the device. In the 2-D system the electrons are confined in the z-direction, while in the 1-D system electrons are confined in the y direction and the z direction.

Silicon has bulk inversion symmetry [11] and thus the Dresselhaus or the bulk inversion asymmetry fields are absent in silicon. Thus the only dominant spin orbit interaction in silicon is the Rashba spin orbit interaction which arises due to structural inversion asymmetry. The transverse field acts as a dominant symmetry breaking electric field and leads to Rashba interaction. Spin depolarization in the silicon channel hence occurs because of Rashba spin orbit coupling.

Thus the D'yakonov-Perel (DP) spin relaxation is accounted by the Rashba interaction and electrons evolve according to the Rashba Hamiltonian [25],

$$H_R^{2D} = \eta(k_y\sigma_x - k_x\sigma_y) \quad (1)$$

$$H_R^{2D} = -\eta k_x\sigma_y \quad (2)$$

where the Rashba coefficient $\eta$ depends on the material and electric field.

Over each free flight time in which no scattering event takes place, the spin vector evolves according to the equation,

$$\frac{d\vec{S}}{dt} = \vec{\Omega} \times \vec{S} \quad (3)$$

where the precession vector $\vec{\Omega}$ has only the Rashba component due to the Rashba Hamiltonian, $\vec{\Omega}_R(k)$ and can be written as,

$$\vec{\Omega}_R(k_x, k_y)^{2D} = \frac{2\eta(k_y\hat{i} - k_x\hat{j})}{\hbar} \quad (4)$$

$$\Omega_R(k_x)^{1D} = -\frac{2\eta k_x\hat{j}}{\hbar} \quad (5)$$

In a 2-D channel, as a result of confinement in the z-direction, the six degenerate valleys split into $\Delta_2$ and $\Delta_4$ valleys. The $\Delta_2$ valleys (valley pairs along z) are lower in energy than the $\Delta_4$ valleys (valley pairs along x and y). This is because the electrons in the perpendicular valleys have electrons with a light conductivity effective mass as compared to the conductivity effective mass of electrons in the longitudinal valleys. The population density of the electrons in $\Delta_4$ valleys is lesser than $\Delta_2$ valleys due to the higher energy of $\Delta_4$ valley. Hence in the 2-D channel, the electrons are assumed to occupy only the $\Delta_2$ valleys and only the $\Delta_2$ valleys are considered for the sake of our simulation.

In a 1-D system, as a result of confinement in the y-direction as well as the z-direction, conductivity effective mass for valley pairs along x is smaller than the conductivity effective mass for valley pairs along *y* and *z*. Due to this, the subband energy levels in valley pairs along *x* are higher than the valley pairs along *y* and *z*. The subbands in the $\Delta_2$ valleys (valley pairs along *x*) get depopulated[26] as compared to the subbands in the $\Delta_4$ valleys (valley pairs along *y* and *z*) Hence in the 1-D nanowires, the electron dynamics is assumed to be limited to the $\Delta_4$ valleys and only $\Delta_4$ valleys are considered in the simulations.

The non-parabolicity of the bands is taken into account [22] by considering energy-wave vector relation of the type,

$$\varepsilon(1+\alpha\varepsilon) = \frac{\hbar^2 k^2}{2m} \qquad (6)$$

where *α* is the non-parabolicity parameter and is given by the following expression,

$$\alpha = \frac{1}{E_g}\left(1 - \frac{m}{m_o}\right)^2 \qquad (7)$$

In Eq.(7) $E_g$ is the bandgap, *m* is the effective mass of the electron and $m_0$ is the rest mass of the electron.

The scattering processes considered are intravalley and intervalley phonon scattering, surface roughness scattering and ionized impurity scattering. Both optical phonons and acoustic phonons have been considered. The effect of subbands is included in the simulation to account for intervalley and intravalley intersubband scatterings.

   a. *Scattering Rates in 1-D nanowire*

In a 1-D system, the scattering rates are taken from Ref.[26,31]. The acoustic phonon scattering is calculated only because of bulk acoustic phonons and the confined acoustic phonons are not considered. The change that comes in the acoustic phonon spectrum because of the acoustic phonon confinement is estimated to be very small [28] and hence is neglected. The bulk acoustic phonon scattering, optical phonon scattering and surface roughness scattering rates have been used from Ref. [26]. The impurity scattering rate has been taken from Lee and Spector's work [27].

   b. *Scattering Rates in 2-D channel*

For the calculation of the scattering rates in a 2-D system we follow the papers of Price [29] and Yokoyama and Hess [30], who formulated the 2D scattering rates for parabolic semiconductors. The acoustic phonon scattering and optical phonon scattering rates have been used from Ref.[31]. The surface roughness and ionized impurity scattering rates have been taken from Ref.[32].

III. RESULTS

The 2-D structure studied is taken to be 5 nm in thickness while the width is 125 nm. The 1-D nanowire structure is taken to be 5 nm both in thickness and width. The doping density is taken to be $5 \times 10^{25}$ /m$^3$. The transverse symmetry breaking effective field is taken to be 100 kV/cm which is a reasonable value for silicon structures. The surface roughness parameters are taken from Ref.[26].

The value of the Rashba coefficient $\eta$ is calculated using the formula [11],

$$\eta = 1.66 \times 10^{-6} \times q\hbar E_{eff} \text{ eV-m} \qquad (8)$$

where $q$ is the electronic charge and $E_{eff}$ is the transverse effective field. For our simulation, $E_{eff}$=100kV/cm which yields $\eta$=1.1 x 10$^{-14}$ eV-m. Four subbands are considered in each valley, both for 1-D and 2-D channels. The moderate values of driving electric field used in our simulations ensures that the majority of electrons are restricted to the first four subbands, thus justifying the use of four subbands in our simulations. In Ref. [8] similarly 3 subbands were considered for the sake of spin polarized transport. The energy levels of subbands are calculated using an infinite potential well approximation. The various other material and scattering parameters are taken to be same as that for bulk silicon and are adopted from a standard book on Monte Carlo simulations [22]. The electrons are injected from the left end (at x=0) with initial polarization along the thickness of the wire i.e. in the z-direction. A time step of 0.2 fs was chosen and electrons were simulated for $8 \times 10^5$ steps so that a steady state is achieved. Data are recorded for the last 40,000 steps only. The ensemble average is computed for spin vector for the last 40,000 steps at each point of the wire.

   a. *Spin Relaxation lengths at room temperature(300K) for a driving electric field of 1kV/cm*

A room temperature (300K) study was done for both 2-D channels and 1-D nanowires at an average field of 1kV/cm. The ensemble averaged magnitude of spin for an initial polarization along the thickness of the wire (in the z-direction) decays along the channel as shown in Fig.1 for 2-D and 1-D channels.

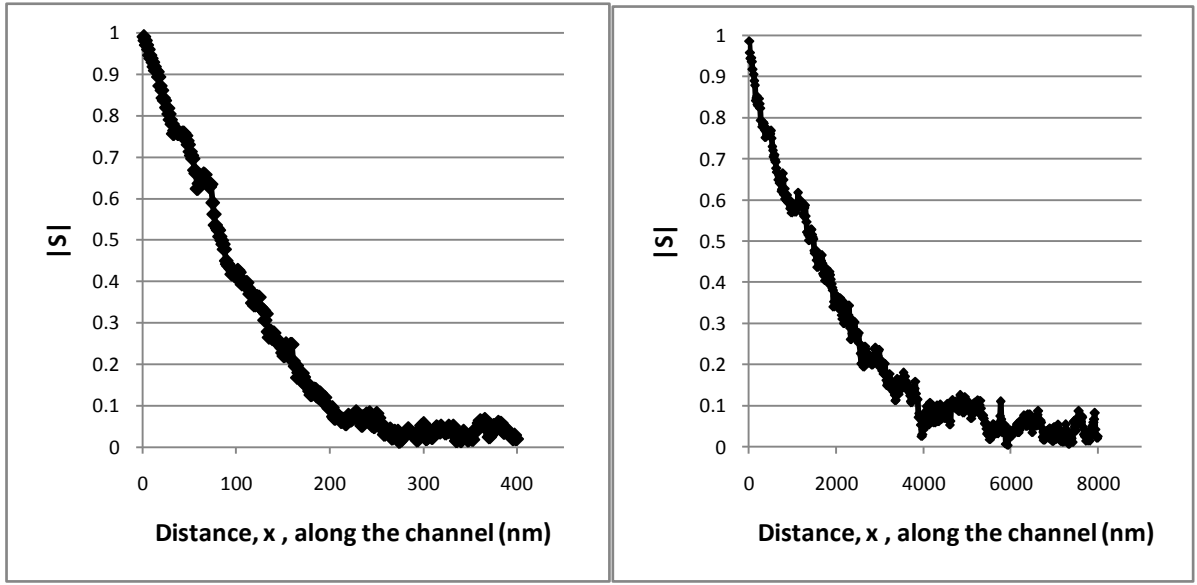

(a)                         (b)

Fig.1: |S| along channel length for electrons in (a) a 2-D channel (b) a 1-D channel.

Spin relaxation length for electrons in 2-D channel is found to be around 125 nm compared to 1.98 μm in a 1-D nanowire. Thus the spin relaxation length for a nanowire is about 16 times larger than two dimensional channels. This result is in conformity with similar studies made by researchers in this field where they have reported similar improvements of atleast an order of magnitude in 1-D channels compared to 2-D channels [13,33,34].

It would seem intuitive to suggest that the difference in scattering rates between nanowire and 2-D channels is the cause for difference in spin relaxations. However this leads to an erroneous interpretation since calculated mobility in 1-D structures have been found lesser than their 2-D counterparts by researchers [26,33]. Usage of the above idea to explain spin relaxation then would suggest that nanowires have smaller spin relaxation lengths which is not what we observe, both experimentally and theoretically. In Fig. 2 scattering rates are plotted against electron energy for both nanowires and 2-D channels and the difference in scattering rates between the two is found to be ever so slight to explain the difference of more than an order of magnitude in spin relaxation lengths.

The origin of this difference in spin relaxation lengths stems from the fact that the dominant spin relaxing mechanism, D-P relaxation is suppressed in a nanowire [35,36]. As pointed out in Ref.[33] and by the Eq.(3), the Rashba interaction causes the electron spin to precess about a precession vector $\vec{\Omega}_R$. In a 1-D system, the electron is constrained to move only in one direction (x-direction in our case) and hence the precession vector always points in a particular direction, as seen from Eq.(5). Any scattering event that changes the velocity of the electron, only ends up changing the magnitude of the precession vector but the not direction which is fixed in space. This leads to slower spin relaxation and hence larger spin relaxation lengths. Now in a 2-D system, the precession vector given by Eq.(4) points along the two direction of motions of electrons (the x and y-directions). A scattering event changes the velocities in both the directions of motion and hence results in a change in both the magnitude and the direction of the precession vector. This leads to a faster randomization of the spin and subsequently smaller spin relaxation lengths. This is consistent with the expectation that more random the motion of electron, faster is the spin relaxation. In 2-D, the electron motion gets randomized along two directions while in 1-D it can happen only in one direction.

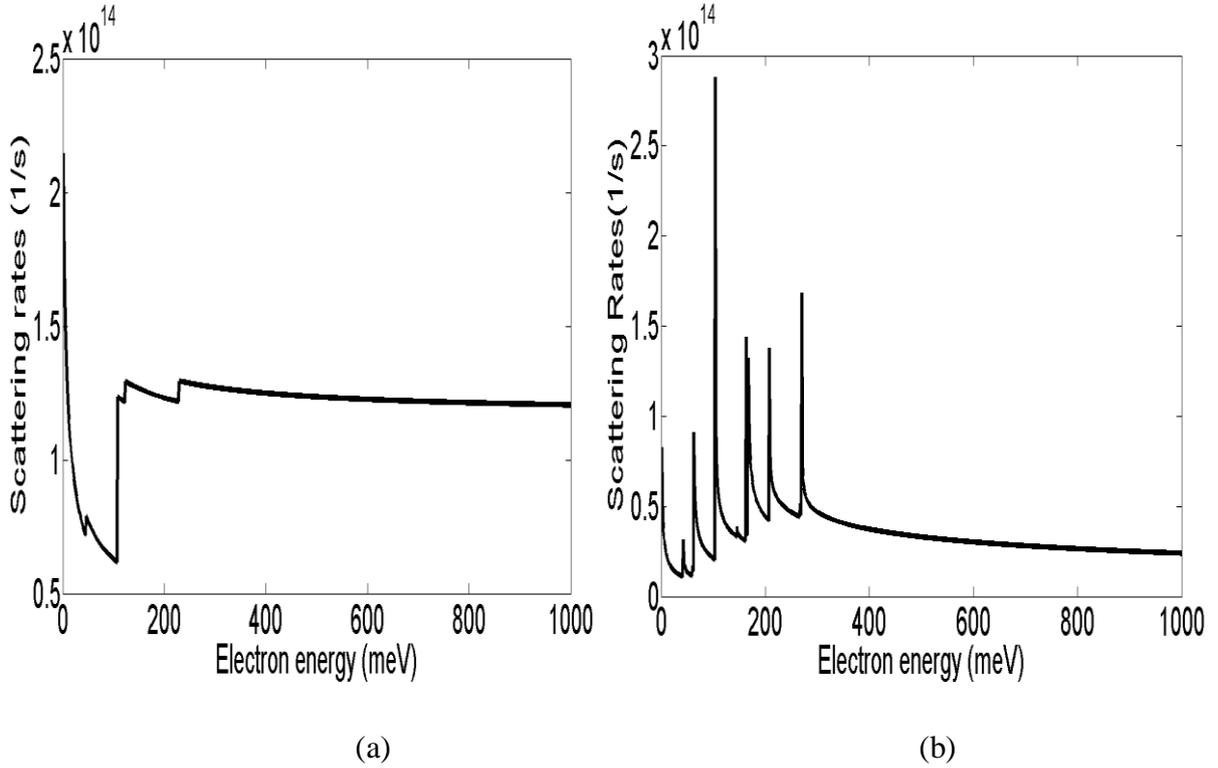

(a)                 (b)

FIG.2: (a) Scattering Rates versus electron energies for 2-D channel (b) Scattering Rates versus electron energies for 1-D nanowire.

### b. *Effect of applied electric field*

The spin relaxation length is expected to be weakly dependent on applied electric field since its overall effect is decided by the individual effects of two competing factors – scattering rates and ensemble averaged drift velocity of the electrons. Any increase in drift velocity over the scattering rate helps the electron and hence the spin to penetrate further into the device and thus will lead to higher spin relaxation lengths. The opposite of this happens when scattering rates dominate over drift velocity and the increased scattering rates will randomize the spin faster reducing the spin relaxation lengths. The overall effect will be decided by the dominant effect amongst the two.

In Fig. 3 decay of the average spin vector at different applied fields is shown for both a 2-D channel and a 1-D nanowire. Fig 3(a) shows the decay of spin along the channel length for a 2-D channel at 100V/cm, 2kV/cm and at 5kV/cm. The spin relaxation length for 2-D channel increases from around 100nm at 100V/cm to120nm at 2kV/cm and to 169nm at 5kV/cm. Fig 3(b) shows the decay of spin along the channel length for a 1-D nanowire at 100V/cm, 2kV/cm and 10kV/cm. The spin relaxation length for 1-D nanowire changes from 2.04µm at 100V/cm to 1.94µm at 2kV/cm and to 2.55µm at 10kV/cm.

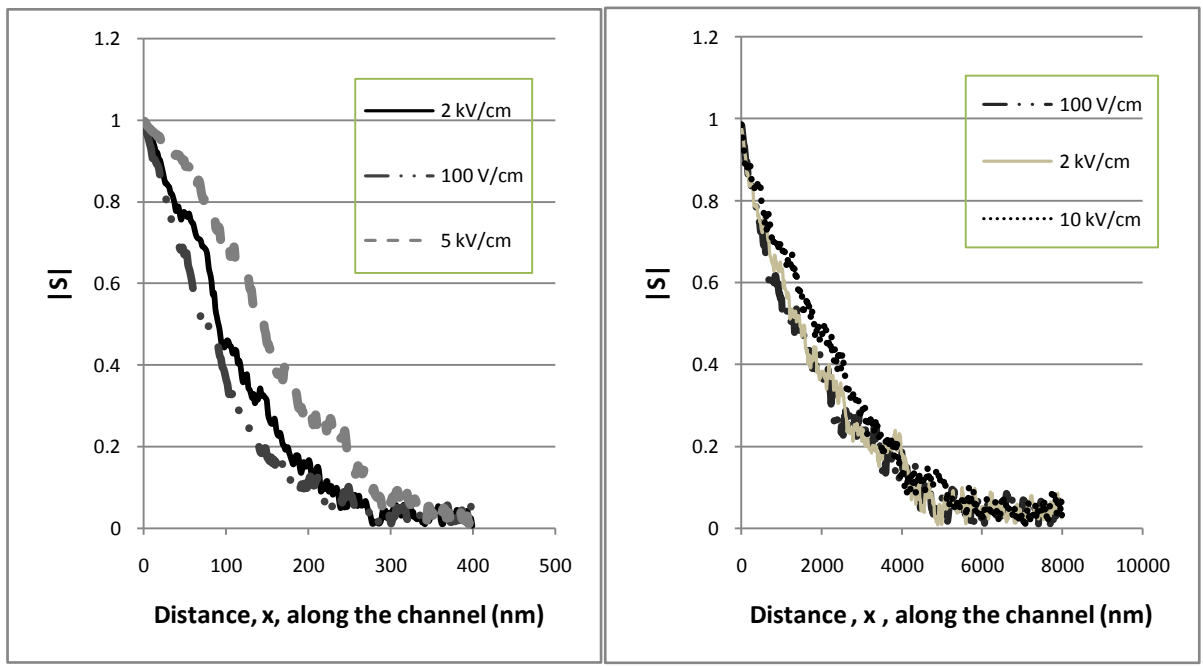

(a) (b)

Fig.3: Variation of |S| with applied electric field. |S| along channel length at 300K for (a) 2-D channel. Solid and dotted line denotes magnitude of spin at an average driving electric field of 100V/cm, solid lines are for 2kV/cm and dashed lines are at a field of 5kV/cm  (b) 1-D nanowire. Thick solid and dotted line denotes magnitude of spin at 100 V/cm, Solid line denotes spin at 1kV/cm and  dotted lines are for 10kV/cm.

Fig. 4 shows the electric field dependence of spin relaxation lengths for both 2-D and 1-D channels. The values of electric field used are moderate which ensures that drift velocity saturation does not occur. The dependence of spin relaxation length with electric field is clearly nonmonotonic. In our four subband model, the intersubband scattering saturates after a point as can be seen from Fig.2. At higher fields and thus at higher electron energies the scattering rates remain fairly constant with only a slight variation. The drift velocity dominates the scattering rates in this regime and spin relaxation length increases as in Fig.4. A local point of minima occurs when the increased scattering rates due to intersubband scattering dominate over the effect due to increase in drift velocity.

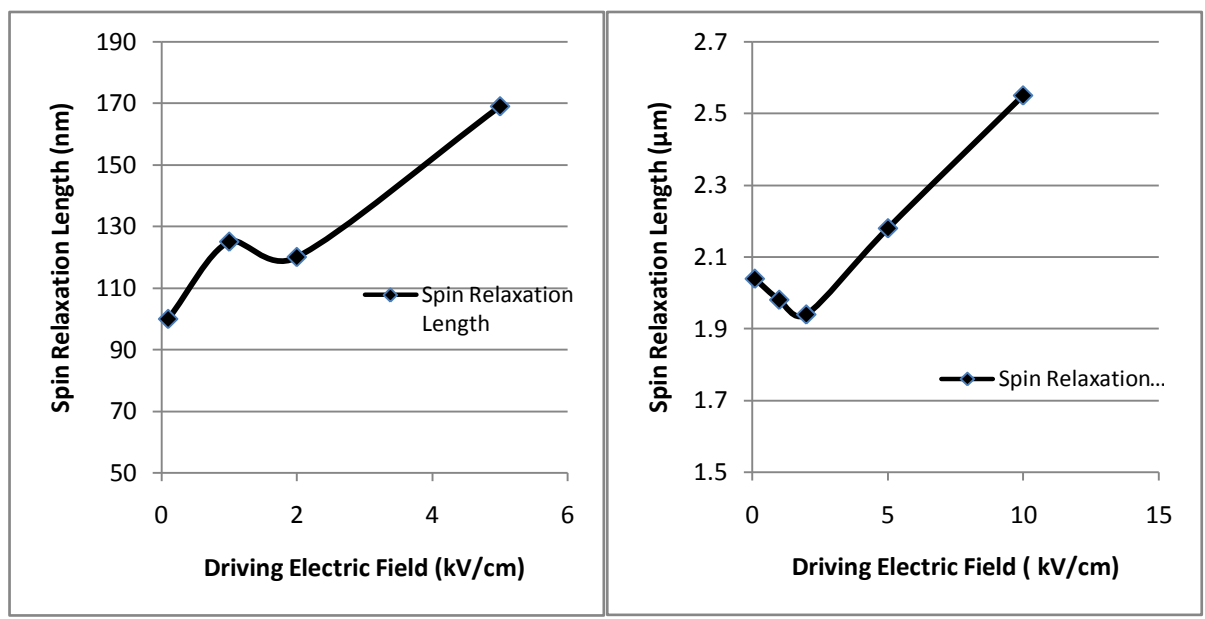

(a) (b)

Fig. 4: Variation of Spin Relaxation length with the driving electric field for (a) 2-D channel (b) 1-D nanowire.

IV. CONCLUSION

Our work shows that at a driving field of 1kV/cm and at room temperature while the 2-D silicon channels have a spin relaxation length of 125nm, confining the motion to only one direction can improve drastically upon the spin relaxation length (more than an order of magnitude to around 1.98μm). This result helps us to draw the conclusion that the spin remains information polarized upto a larger length on using 1-D channels. This larger spin relaxation length is significantly important for spin based information devices as many spin based devices can be best implemented with nanowire channels as opposed to 2-D channels. Our work suggests that silicon can be employed in spintronic related applications and we hope that experimental studies on spin transport in silicon to determine the spin relaxation length will be taken up to validate our claim. We also intend to undertake such spin transport studies on different materials of potential interest in the future.